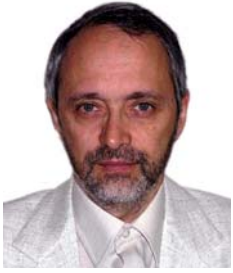

# THERMOELECTRIC PROPERTIES OF MACROSCOPICALLY INHOMOGENEOUS COMPOSITES

A.A.Snarskii[1], I.V.Bezsudnov[2]
([1] "KPI" National Technical University of Ukraine, Kyiv, Ukraine;
[2] "Nauka-Service" Ltd, Moscow, Russia)

*A.A. Snarskii*

*I.V. Bezsudnov*

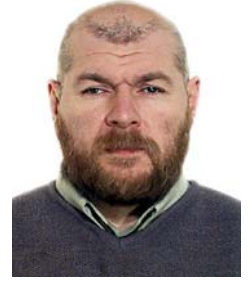

- *Review of theoretical methods for description of thermoelectrical properties of two-phase macroscopically inhomogeneous media is given. Description by means of a hierarchical model of percolation structure is given, allowing a description on a qualitative and quantitative level of the basic regularities of effective thermoelectric coefficient close to flow threshold.*

**Introduction**

Thermoelectric effects exemplify physical phenomena when two thermodynamic flows simultaneously (electric flow density $\mathbf{j}$, heat flow density $\mathbf{q}$) are created by two thermodynamic forces: electric field intensity $\mathbf{E}$ and temperature gradient $\nabla T$

$$\begin{cases} \mathbf{j} = \sigma \mathbf{E} + \sigma \alpha (-\nabla T), \\ \dfrac{\mathbf{q}}{T} = \sigma \alpha \mathbf{E} + \kappa (1 + ZT)(-\nabla T), \quad ZT = \dfrac{\sigma \alpha^2}{\kappa} T, \end{cases} \quad (1)$$

where $\alpha$ is thermoelectric coefficient, $\kappa$ is thermal conductivity, and $Z$ is figure of merit (Ioffe's number).

Thermoelectric effects enter in (1) the cross components, owing to the existence of which the electric current is induced not only by electric field, but also by temperature gradient, and, respectively, part of heat flow is related to electric field (or, what is the same, to electric current, this being the essence of the Peltier effect).

The properties of a macroscopically inhomogeneous medium are described on the whole by the so-called effective kinetic coefficients. By definition, the effective kinetic coefficients interconnect average-volume thermodynamic flows and forces (see, for example, review [1]). In particular, from (1) for effective electric conductivity $\sigma_e$ and thermoelectric coefficient $\alpha_e$ it follows that $\langle \mathbf{j} \rangle = \sigma_e \langle \mathbf{E} \rangle + \sigma_e \alpha_e \langle -\nabla T \rangle$, where $\langle ... \rangle$ means averaging with respect to volume.

Just as the problem of calculating effective conductivity, the problem of calculating effective kinetic coefficient (including the problem of finding effective thermoelectric coefficient) can be solved in various approximations. Consider first the EMT-approximation.

**1. EMT-approximation**

To construct the EMT-approximation (abbreviation EMT originates from Effective Medium Approximation – one of the variants of medium field theory or self-consistent field theory [2, 3]), for thermoelectric effects it is necessary, like in the $\sigma_e$ problem, to have solutions of problem on the distribution of fields and flows for a solitary spherical inclusion of one phase into another. In the case being considered this inclusion consisting of material with coefficients $\sigma_1$, $\kappa_1$, $\alpha_1$ (or $\sigma_2$, $\kappa_2$, $\alpha_2$) is in material with coefficients $\sigma_e$, $\kappa_e$, $\alpha_e$, and the values of field $\langle \mathbf{E} \rangle$ and temperature gradient $\langle \nabla T \rangle$ are assigned on the infinity. Equations allowing to find potential distribution $\varphi(\mathbf{r})$ and temperature distribution $T(\mathbf{r})$ for the general case are given, for example, in [4]. In [5], in the approximation when corrections for $\sigma_e$ and $\kappa_e$ are small due to thermoelectric effects, the analytical solution of this problem is given, from whence follows the expression for $\alpha_e$

$$\alpha_e = \frac{\langle \alpha\sigma/\Delta_0 \rangle}{\langle \sigma/\Delta_0 \rangle}, \quad \Delta_0 = (2\sigma_e + \sigma)(2\kappa_e + \kappa), \tag{2}$$

where $\sigma_e$ and $\kappa_e$ are determined from the EMT-approximation (without taking into consideration thermoelectric effects)

$$\left\langle \frac{\sigma_e - \sigma}{2\sigma_e + \sigma} \right\rangle = 0, \quad \left\langle \frac{\kappa_e - \kappa}{2\kappa_e + \kappa} \right\rangle = 0. \tag{3}$$

For a two-phase medium (in a three-dimensional case) from (2) and (3) it follows:

$$\alpha_e = \frac{p\sigma_1\alpha_1(2\sigma_e + \sigma_2)(2\kappa_e + \kappa_2) + (1-p)\sigma_2\alpha_2(2\sigma_e + \sigma_1)(2\kappa_e + \kappa_1)}{p\sigma_1(2\sigma_e + \sigma_2)(2\kappa_e + \kappa_2) + (1-p)\sigma_2(2\sigma_e + \sigma_1)(2\kappa_e + \kappa_1)}. \tag{4}$$

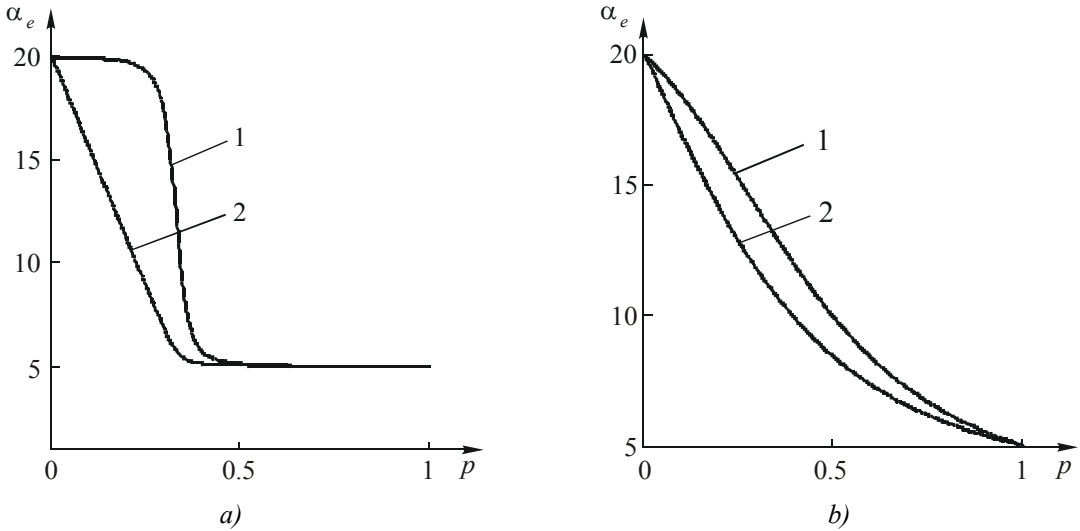

*Fig.1. Concentration dependence $\alpha_e = \alpha_e(p)$ for two limiting cases of phase thermal conductivity ratio $\kappa_2/\kappa_1 \approx 1$ and $\kappa_2/\kappa_1 \ll 1$ – EMT-approximation. Numerical values of coefficients here and afterwards are selected as arbitrary values.*
*a) Strong inhomogeneity of conductivity $\sigma_1 \gg \sigma_2$, as an example, selected $\sigma_1/\sigma_2 = 10^3$, $\alpha_2/\alpha_1 = 4$. Stepped dependence 1 is matched with strong inhomogeneity of phase thermal conductivity $\kappa_2/\kappa_1 = 5\cdot10^{-2} \ll 1$, dependence 2 (step is absent) with $\kappa_1 = \kappa_2$.*
*b) In case of weak inhomogeneity of phase conductivity $\sigma_1/\sigma_2 = 6$ stepped dependence disappears, 1 – $\kappa_2/\kappa_1 \ll 1$, 2 – $\kappa_1 = \kappa_2$.*

For a two-dimensional case instead of (3) one should use the relationships $\left\langle \frac{\sigma_e - \sigma}{\sigma_e + \sigma} \right\rangle = 0$, $\left\langle \frac{\kappa_e - \kappa}{\kappa_e + \kappa} \right\rangle = 0$ and $\Delta_0 = (\sigma_e + \sigma)(\kappa_e + \kappa)$. Then [4]

$$\alpha_e = \frac{p\sigma_1\alpha_1(2\sigma_e + \sigma_2)(\kappa_e + \kappa_2) + (1-p)\sigma_2\alpha_2(\sigma_e + \sigma_1)(\kappa_e + \kappa_1)}{p\sigma_1(2\sigma_e + \sigma_2)(\kappa_e + \kappa_2) + (1-p)\sigma_2(\sigma_e + \sigma_1)(\kappa_e + \kappa_1)} \quad (5)$$

From analysis (4) it follows that concentration dependence $\alpha_e = \alpha_e(p)$ will be different, depending on phase thermal conductivity ratio $\kappa_2/\kappa_1$ (Fig.1).

With a large inhomogeneity of electric conductivity $\sigma_1/\sigma_2 \gg 1$ and thermal conductivity $\kappa_1/\kappa_2 \gg 1$, in the concentration dependence of effective thermoelectric coefficient $\alpha_e(p)$ a step can be observed that disappears at $\kappa_1 \approx \kappa_2$. Dependence shown in Fig.1 on the right is observed with a weak inhomogeneity of conductivity.

When there is no current flow in one of the phases, i.e., when $\sigma_2 = 0$, in the entire area of $\sigma_e$ existence, i.e. at $p > p_c$, the effective thermoelectric coefficient is constant $\alpha_e = \alpha_1$ and equal to pure phase thermoEMF value (Fig.2).

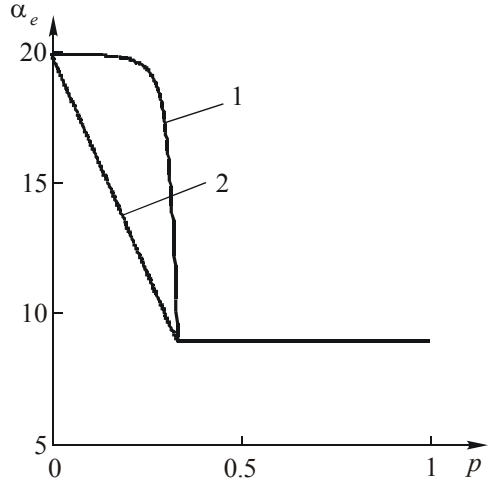

Fig.2. Concentration dependence $\alpha_e = \alpha_e(p)$ at $\sigma_2 = 0$ for two limiting cases
of phase thermal conductivity ratio; $1 - \kappa_2/\kappa_1 \ll 1$ and $2 - \kappa_1 = \kappa_2$.

In the case of $\kappa_1 \gg \kappa_2$ a sharp (percolation) transition of effective thermoelectric coefficient $\alpha_e$ from $\alpha_1$ to $\alpha_2$ is easily observable near $p_c$.

For the general case, without requirement of $ZT \ll 1$, a system of equations to determine $\alpha_e$, $\sigma_e$ and $\kappa_e$ was obtained (within EMT) in [4], however, its analytical investigation is difficult.

## 2. Self-dual media

Like in the "ordinary" (without thermoelectric phenomena) case of inhomogeneous medium, precise solution of the problem on effective thermoelectric coefficient $\alpha_e$ in a 2-dimensional case for self-dual media is possible. There is an infinite and quite versatile set of two-dimensional structures with half concentration of phases, for which the effective conductivity has the same expression, there being no restriction for inhomogeneity value, including the case $\sigma_1/\sigma_2 \to \infty$. These are the so-called self-dual media ($D$-media). For the first time the expression for $\sigma_e$ in $D$-medium was obtained in [6] for a very general case. Symmetry transformations for local fields and currents, generalized and used later on for solving numerous problems, were introduced in the general form in the same paper. Consider a two-dimensional two-phase medium with such arrangement of phases when:

– effective conductivity is isotropic,

– $\sigma_e$ is not varied due to interchange of phases.

The latter means that phases are in geometrically equivalent positions and their concentration is equal to 0.5.

If for two-dimensional randomly inhomogeneous media close to flow threshold on larger than correlation length dimensions various phases are also in geometrically equivalent (medium) position, the interchange of $\sigma_1 \Leftrightarrow \sigma_2$ will not vary $\sigma_e$. Thus, these are also $D$-media.

For the effective thermoelectric coefficient $\alpha_e$ of self-dual media in [4, 7] it was obtained

$$\alpha_e = \frac{\alpha_1\sqrt{\sigma_1\kappa_2} + \alpha_2\sqrt{\sigma_2\kappa_1}}{\sqrt{\sigma_1\kappa_2} + \sqrt{\sigma_2\kappa_1}} = \frac{\alpha_1\sqrt{\frac{\kappa_2}{\kappa_1}} + \alpha_2\sqrt{\frac{\sigma_2}{\sigma_1}}}{\sqrt{\frac{\sigma_2}{\sigma_1}} + \sqrt{\frac{\kappa_2}{\kappa_1}}}. \qquad (6)$$

Specifically, at $\sigma_2 = 0$ $\alpha_e(\sigma_2 = 0) = \alpha_1$ (Fig.2). Effective coefficients $\sigma_e$ and $\kappa_e$ for self-dual media are of the form:

$$\sigma_e = \sqrt{\sigma_1\sigma_2} \frac{\sqrt{\sigma_1\kappa_2} + \sqrt{\sigma_2\kappa_1}}{\sqrt{\left(\sqrt{\sigma_1\kappa_2} + \sqrt{\sigma_2\kappa_1}\right)^2 + T\sigma_1\sigma_2(\alpha_1 - \alpha_2)^2}}, \qquad (7)$$

$$\kappa_e = \sqrt{\kappa_1\kappa_2} \frac{\sqrt{\sigma_1\sigma_2}}{\sigma_e}. \qquad (8)$$

As long as flow threshold in self-dual media $p_c = 1/2$, expression (6) gives the value of effective thermoelectric coefficient $\alpha_e$ at flow threshold in the two-dimensional case. At close phase thermal conductivities $\kappa_1 \approx \kappa_2$ and strong inhomogeneity of conductivity $\sigma_2/\sigma_1 \ll 1$, from (6) it follows

$$\alpha_e = \frac{\alpha_1 + \alpha_2\sqrt{h}}{1+\sqrt{h}} \approx \alpha_1 + \alpha_2\sqrt{h}, h = \sigma_2/\sigma_1 \ll 1. \qquad (9)$$

As a rule, in "metal" (with $\sigma_1$ conductivity) phase thermoEMF $\alpha_1$ is much lower than in "semiconductor" $\alpha_1 \ll \alpha_2$, then from (9) near behaviour threshold, if $\alpha_2\sqrt{h} \gg \alpha_1$

$$\alpha_e \approx \alpha_2 \sqrt{h}, \qquad (10)$$

i.e. effective thermoelectric coefficient $\alpha_e$ is limited by a small factor $h = \sigma_2/\sigma_1$.

At $\kappa_2 \ll \kappa_1$ and, as before, $\sigma_2/\sigma_1 \ll 1$ and $\alpha_1 \ll \alpha_2$ from (4) it follows

$$\alpha_e \approx \alpha_2 \frac{\sqrt{\sigma_2/\sigma_1}}{\sqrt{\sigma_2/\sigma_1} + \sqrt{\kappa_2/\kappa_1}}, \qquad (11)$$

and, if $\sqrt{\sigma_2/\sigma_1} \approx \sqrt{\kappa_2/\kappa_1}$

$$\alpha_e \approx \alpha_2, \qquad (12)$$

i.e. as compared to the case $\kappa_1 \approx \kappa_2$, at $p = p_c$ the effective thermoelectric coefficient $\alpha_e$ rises steeply. This behaviour agrees well with the results of EMT-approximation (Fig.1a).

If (1) is written in the abstract form, suitable for any two-flow system with cross phenomena

$$\begin{aligned} \mathbf{j} &= A_{11}\mathbf{e} + A_{12}\mathbf{g}, \\ \mathbf{q} &= A_{21}\mathbf{e} + A_{22}\mathbf{g}, \end{aligned} \quad \hat{A} = \begin{pmatrix} A_{11} & A_{12} \\ A_{21} & A_{22} \end{pmatrix}, \qquad (13)$$

where $\hat{A}$ is a tensor of local kinetic coefficients, the effective values $A_{i,j}^e$ of self-dual media (with the accuracy of notation coinciding with (6) – (8)) will be of a simple and compact form [1]

$$A^e = \left(Det\hat{A}_1\, Det\hat{A}_2\right)^{\frac{1}{4}} \frac{\hat{\Omega}_1 + \hat{\Omega}_2}{\sqrt{Det\left(\hat{\Omega}_1 + \hat{\Omega}_2\right)}}, \quad \hat{\Omega}_n = \frac{\hat{A}_n}{\sqrt{Det\left(\hat{A}_n\right)}}, \quad n = 1, 2, \qquad (14)$$

where the subscript $n$ denotes phase number.

Just as for the problem on effective conductivity, there is an exact solution for a two-dimensional polycrystal [6]

$$A^e = \left(Det\hat{A}_\parallel Det\hat{A}_\perp\right)^{\frac{1}{4}} \frac{\hat{\Omega}_\parallel + \hat{\Omega}_\perp}{\sqrt{Det\left(\hat{\Omega}_\parallel + \hat{\Omega}_\perp\right)}}, \quad \hat{\Omega}_{\parallel,\perp} = \frac{\hat{A}_{\parallel,\perp}}{\sqrt{Det\left(\hat{A}_{\parallel,\perp}\right)}}, \qquad (15)$$

where $\hat{A}_\parallel$ and $\hat{A}_\perp$ are principal values of tensor of local coefficients $A_{ik}$.

In "thermoelectric" terms for a two-dimensional polycrystal

$$\alpha^e = \frac{\alpha_\parallel \sqrt{\sigma_\parallel \kappa_\perp} + \alpha_\perp \sqrt{\sigma_\perp \kappa_\parallel}}{\sqrt{\sigma_\parallel \kappa_\perp} + \sqrt{\sigma_\perp \kappa_\parallel}}, \qquad (16)$$

$$\sigma^e = \sqrt{\sigma_\parallel \sigma_\perp} \frac{\sqrt{\sigma_\parallel \kappa_\perp} + \sqrt{\sigma_\perp \kappa_\parallel}}{\sqrt{\left(\sqrt{\sigma_\parallel \kappa_\perp} + \sqrt{\sigma_\perp \kappa_\parallel}\right)^2 + T\sigma_\parallel \sigma_\perp \left(\alpha_\parallel - \alpha_\perp\right)^2}}, \qquad (17)$$

$$\kappa^e = \sqrt{\kappa_\parallel \kappa_\perp} \frac{\sqrt{\sigma_\parallel \sigma_\perp}}{\sigma_e}, \qquad (18)$$

where $\sigma_\parallel, \kappa_\parallel, \alpha_\parallel$ and $\sigma_\perp, \kappa_\perp, \alpha_\perp$ are diagonal components of electric conductivity, thermal conductivity and thermoEMF tensors.

The expressions for $\sigma_e$ and $\kappa_e$ (7) can be written such as to explicitly single out a component with "thermoelectric figure of merit"

$$\sigma_e = \frac{\sqrt{\sigma_1 \sigma_2}}{\sqrt{1+ZT}}, \qquad (19)$$

where under $ZT$ in this case the following value is meant:

$$ZT = \frac{\sqrt{\sigma_1 \sigma_2}}{\sqrt{\kappa_1 \kappa_2}} \frac{(\alpha_1 - \alpha_2)^2}{\left( \sqrt{\frac{\sigma_2}{\sigma_1}} \Big/ \sqrt{\frac{\kappa_2}{\kappa_1}} + \sqrt{\frac{\kappa_2}{\kappa_1}} \Big/ \sqrt{\frac{\sigma_2}{\sigma_1}} \right)^2} T, \qquad (20)$$

$$\kappa_e = \sqrt{\kappa_1 \kappa_2} \sqrt{1+ZT}. \qquad (21)$$

Thus, condition of small effect of thermoelectric phenomena on effective electric and thermal conductivities is immediately obvious: $ZT \ll 1$. The concept of thermoelectric figure of merit (Ioffe's number) $Z_i = \sigma_i \alpha_i^2 / \kappa_i$ is determining for the calculation of the efficiency of thermoelectric devices (thermal generators, Peltier coolers, etc.) [8, 9]. Note that in $ZT$ (20) parameters of thermoelectric materials $\sigma_1, \kappa_1, \alpha_1$ and $\sigma_2, \kappa_2, \alpha_2$ form a peculiar combination (just as $Z = \sigma \alpha^2 / \kappa$), which is not, however, a direct function of $Z_1 = \sigma_1 \alpha_1^2 / \kappa_1$ and $Z_2 = \sigma_2 \alpha_2^2 / \kappa_2$. It means, at least in this case, that the influence of thermoelectric phenomena on the effective composite properties (including the efficiency of thermoelectric devices made of this composite) cannot be expressed only through $Z_1$ and $Z_2$ (or through $Z_e = \sigma_e \alpha_e^2 / \kappa_e$), but calls for more sophisticated (nontrivial) combinations of parameters, such, for example, as in (20).

Double-sided limitations of $Z_e$ were dealt with in [10-12].

## 3. Critical concentration area – behaviour of $\alpha_e$ close to flow threshold

Now consider behaviour of effective thermoelectric coefficient close to flow threshold $p_c$, i.e. when proximity to flow threshold $\tau = (p - p_c)/p_c$ is much less than unity in absolute value. The concentration value $|\tau| \ll 1$ is called a critical value. In a medium with strongly inhomogeneous conductivity ($h = \sigma_2/\sigma_1 \ll 1$) the distribution of currents and fields in a medium close to flow threshold, i.e. at $|\tau| \ll 1$ is "controlled" by percolation structure – bridges and interlayers [13-15]. The presence of additional processes, in this case thermoelectric, certainly, affects the distribution of currents in a medium, however, bridges and interlayers remain the governing elements of percolation structure, as before. Therefore, it should be expected (and these expectations are justified) that behaviour of effective kinetic coefficients, including $\alpha_e$, close to flow threshold, will be universal, i.e. will be described by critical indexes.

Research on behaviour of $\sigma_e$, $\kappa_e$ and $\alpha_e$ in critical area started in [16-17]. In [18-19] numerical experiments were performed to determine critical indexes. Prior to discussing these and other papers, let us explain on a simple good-quality physical level the critical behaviour of $\alpha_e$ for two different cases: $\kappa_1 \gg \kappa_2$ and $\kappa_1 \approx \kappa_2$. In so doing, we will consider that $\sigma_2/\sigma_1 \ll 1$ and $\alpha_2 \gg \alpha_1$. As it will be obvious, a hierarchical model of percolation structure [13-15] makes possible not only to show the difference in behaviour of $\alpha_e$ for these cases, but also to determine critical indexes $\alpha_e$.

Consider first the medium below flow threshold [20]. In the presence of temperature gradient this scheme of hierarchical model is of the form (Fig.3).

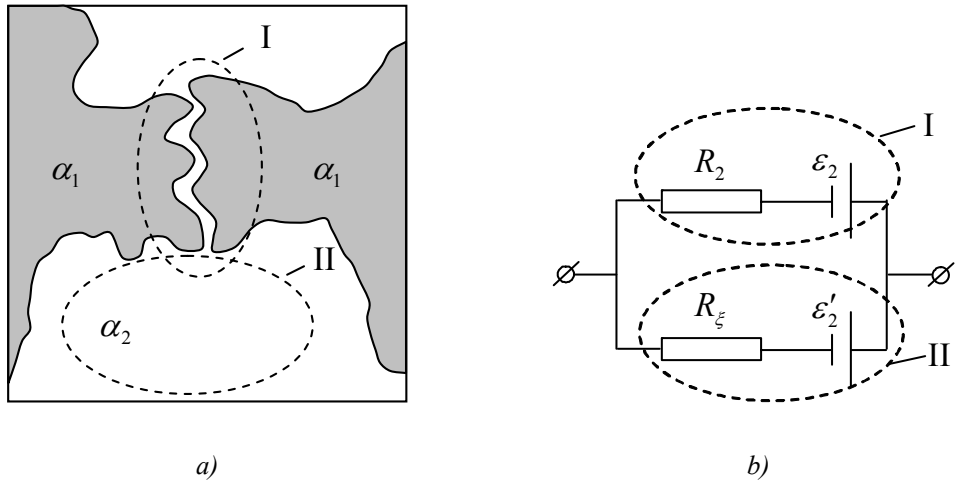

*a)* *b)*

*Fig.3. Percolation structure below flow threshold (a) and its equivalent electric scheme (b) with regard for EMF originated due to temperature difference*
$\Delta T = T_2 - T_1$. $R_2$ – *interlayer resistance;*
$\varepsilon_2$ – *EMF originated in the interlayer;* $R_\xi$ – *resistance of parallel to interlayer portion of poorly conducting medium (with characteristic dimension $\xi$);*
$\varepsilon_2'$ – *EMF originated in this portion.*

Denoting temperature difference on dimensions $\xi$ as $\Delta T = T_2 - T_1$, the total EMF $E_\xi$ on dimensions $\xi$ (scheme 3b) will be found as

$$E_\xi = \left( \frac{E_2}{R_2} + \frac{E_2'}{R_\xi} \right) \bigg/ \left( \frac{1}{R_2} + \frac{1}{R_\xi} \right), \qquad (22)$$

whence

$$\alpha_e = \frac{E_\xi}{\Delta T} = \frac{E_2 + E_2' \frac{R_2}{R_\xi}}{1 + \frac{R_2}{R_\xi}} \cdot \frac{1}{\Delta T} \approx \frac{E_2 + E_2' \frac{R_2}{R_\xi}}{\Delta T}, \quad (23)$$

where it is taken into account that interlayer resistance $R_2$ is always much lower than $R_\xi$ – the resistance of that part of poorly conducting phase which is not included in the interlayer, and

$$E_2 = \Delta\alpha_2 \Delta T, \quad E_2' = \Delta\alpha_2 \Delta T, \quad \Delta\alpha = \alpha_2 - \alpha_1, \quad (24)$$

where $\Delta T_2$ is temperature difference on the interlayer.

With a large difference in phase thermal conductivities $\kappa_1 \gg \kappa_2$, temperature difference on the interlayer is practically the same as that on the bulk boundaries $\Delta T_2 \approx \Delta T$, then, on substituting (24) into (23), we get

$$\alpha_e \approx \Delta\alpha \tau^0, \quad \kappa_1 \gg \kappa_2, \quad (25)$$

i.e. in the case $p < p_c$, $\kappa_1 \gg \kappa_2$, $\sigma_1 \gg \sigma_2$, $\alpha_2 \gg \alpha_1$ $\alpha_e$ is independent of concentration.

However, if phase thermal conductivities are about equal – $\kappa_1 \approx \kappa_2$, temperature difference on the interlayer is much less than $\Delta T$

$$\Delta T_2 \approx \Delta T \frac{a_0}{\xi}, \quad \Delta T |\tau|^\nu \ll \Delta T, \quad (26)$$

and then from (23)

$$\alpha_e = \frac{\Delta T\, a_0/\xi + \Delta T\, R_2/R_\xi}{\Delta T} \approx \Delta\alpha \left( |\tau|^\nu + |\tau|^q \right), \quad (27)$$

where it is taken into account that $a_0/\xi \approx |\tau|^\nu$, and $R_2/R_\xi \approx |\tau|^q$ (see the hierarchical model of percolation structure [13]).

Since both in two- and three-dimensional cases $\nu > q$, then $|\tau|^q > |\tau|^\nu$ and from (27) it follows:

$$\alpha_e \approx \Delta\alpha |\tau|^q, \quad \kappa_1 \approx \kappa_2. \quad (28)$$

Weak dependence of effective thermoelectric coefficient $\alpha_e$ on $\tau$ at $\kappa_1 \approx \kappa_2$ and power reduction of $\alpha_e$ at $\kappa_1 \gg \kappa_2$ is predicted by EMT-approximation.

Above flow threshold, percolation structure (according to hierarchical model of percolation structure) is reduced to the following equivalent scheme (Fig.4):

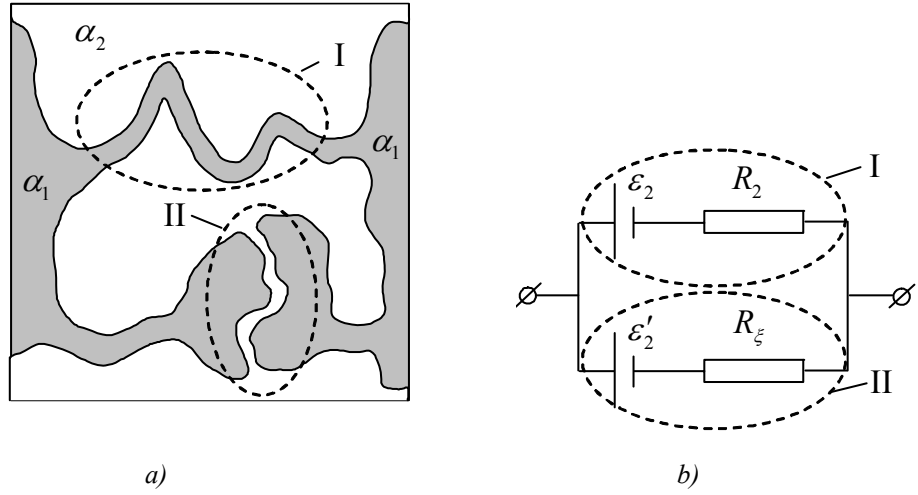

*Fig.4.*
*a) percolation structure above flow threshold;*
*b) its equivalent electric scheme, with regard for EMF, originating due to temperature difference* $\Delta T = T_2 - T_1$.
*$R_1$ and $R_2$ – resistance of bridge and interlayer, $E_1$ and $E_2$ – EMF originating due to temperature difference in the bridge $\Delta T_1$ and $\Delta T_2$.*

Similarly to (23) for effective thermoelectric coefficient $\alpha_e$, according to scheme in Fig. 4b

$$\alpha_e = \frac{E_1/R_1 + E_2/R_2}{1/R_1 + 1/R_2} \cdot \frac{1}{\Delta T} = \frac{E_2 + R_2/R_1}{1 + R_2/R_1} \cdot \frac{1}{\Delta T}. \tag{29}$$

In the case of a big difference between phase thermal conductivities $\kappa_1 \gg \kappa_2$, temperature differences on the bridge and interlayer are about equal

$$\Delta T_1 \approx \Delta T, \ \Delta T_2 \approx \Delta T. \tag{30}$$

For the bridge it is related to the fact that bridge dimension (the distance on a straight line from one end to the other) due to meandering is less than its length, and, as follows from simple geometrical considerations, is proportional to $\xi$.

According to (30)

$$E_1 \approx \alpha_1 \Delta T, \ E_2 \approx \alpha_2 \Delta T, \tag{31}$$

which, on substitution to (29), with regard for $R_2/R_1 = G_2/G_1 = (\sigma_2/\sigma_1) \cdot |\tau|^{t+q}$ (according to hierarchical model of percolation structure) results in

$$\alpha_e \approx \frac{\Delta\alpha + \alpha_1(\sigma_1/\sigma_2)|\tau|^{t+q}}{1 + (\sigma_1/\sigma_2)|\tau|^{t+q}}. \tag{32}$$

Taking into account in (32) the fact that above the smeared region $|\tau|^{t+q} \gg \sigma_1/\sigma_2$ gives

$$\alpha_e \approx \alpha_1 + \alpha_2 \frac{\sigma_2}{\sigma_1} |\tau|^{-(t+q)}, \qquad (33)$$

and then it becomes clearly obvious that for effective thermoelectric coefficient two different behaviour types are possible, depending on the ratio $(\alpha_2/\alpha_1)(\sigma_2/\sigma_1)$ and $|\tau|^{-(t+q)}$.

At

$$|\tau|^{t+q} \gg \frac{\alpha_2}{\alpha_1}\frac{\sigma_2}{\sigma_1}, \qquad (34)$$

from (33) it follows that effective thermoelectric coefficient is practically independent of the concentration

$$\alpha_e \approx \alpha_1 \qquad (35)$$

and is determined by the value of thermoEMF of the first phase.

On approaching flow threshold, i.e. with decreasing $\tau$, when the inverse to (34) inequality holds true, but on condition that the system still remains beyond the smeared region

$$\frac{\sigma_2}{\sigma_1} \ll |\tau|^{t+q} \ll \frac{\alpha_2}{\alpha_1}\frac{\sigma_2}{\sigma_1}, \qquad (36)$$

from (33) it follows

$$\alpha_e \approx \alpha_2 \frac{\sigma_2}{\sigma_1}|\tau|^{-(t+q)}. \qquad (37)$$

These and certain other regularities of $\alpha_e$ behaviour were also obtained by rigorous methods (see, for example, [21-22]), the results of which will be described below. The use of hierarchical model of percolation structure, created on the basis of information on behaviour of $\sigma_e$ close to flow threshold, provides a vivid means for obtaining and explaining the nontrivial peculiarities in behaviour of effective thermoelectric coefficient $\alpha_e$.

A method, proposed in [16] and afterwards generalized and developed in many papers (see, for example, [11, 17, 23-24]), makes it possible in some cases to reduce a two-flow problem with a cross component (with thermoelectric problem as a partial case thereof) to a one-flow problem − for example, the problem of conductivity. That is, this method allows establishing a strict correspondence (isomorphism) between the problem of finding effective kinetic coefficient (EKC) for a system with thermoelectric phenomena and that of finding effective electric conductivity in a medium without thermoelectric phenomena.

Apparently, the simplest illustration of reducing one problem to another belongs to A.M.Dykhne [25]. Specifically, in [25] it is shown how to extend the result [6] $\sigma_e = \sqrt{\sigma_1\sigma_2}$ to thermoelectric case. Let us write equations (1) as

$$\left.\begin{array}{l}\mathbf{j} = s\mathbf{e} + a\mathbf{h}\\ \mathbf{q} = b\mathbf{e} + k\mathbf{h}\end{array}\right\}, \qquad (38)$$

where, in the case of thermoelectric problem $\mathbf{e} = -\nabla\varphi$, $\mathbf{h} = -\nabla T$ and $div\mathbf{j} = 0$, $div\mathbf{q} = 0$.

Local coefficients $s$, $a$, $b$, $k$ are constant in each of the phases and equal to $s_1$, $a_1$ ... and $s_2$, $a_2$ ... in the first and the second phase, respectively.

Let us introduce a new flow $\mathbf{i}$

$$\mathbf{i} = \mathbf{j} + c\mathbf{q} = (s+cb)\mathbf{e} + (a+ck)\mathbf{h} \tag{39}$$

where $c$ is certain coordinate-independent constant, to be selected so as

$$\frac{a_1 + ck_1}{s_1 + cb_1} = \frac{a_2 + ck_2}{s_2 + cb_2} = \omega, \ \omega = \text{const}. \tag{40}$$

Then (39), with regard to (40), can be rewritten as

$$\mathbf{i} = (s+cb)\boldsymbol{\varepsilon}, \tag{41}$$

where $\boldsymbol{\varepsilon}$ is a new field

$$\boldsymbol{\varepsilon} = \mathbf{e} + \omega\mathbf{h} \tag{42}$$

Field $\mathbf{E}$ and flow $\mathbf{j}$ satisfy the same equations $div\mathbf{j} = 0$, $rot\,\boldsymbol{\varepsilon} = 0$, as the initial $\mathbf{j}$, $\mathbf{q}$ and $\mathbf{e}$, $\mathbf{h}$. Therefore, average "current" $<\mathbf{i}> = \mathbf{I}$ and average field $<\boldsymbol{\varepsilon}> = <\mathbf{e}> + \omega<\mathbf{h}>$ are related in the same way as average field and average current in the one-flow problem. If, for example, in the one-flow problem we select the case of a self-dual medium for which $<\mathbf{j}> = \sqrt{\sigma_1\sigma_2}<\mathbf{e}>$, then

$$\mathbf{I} = \sqrt{(s_1+cb_1)(s_2+cb_2)}(<\mathbf{e}> + \omega<\mathbf{h}>). \tag{43}$$

Equation (40) for determination of $c$ is quadratic, its solution yields two values $c_I$ and $c_{II}$. Thus, there are two equations

$$\left.\begin{array}{l}<\mathbf{j}> + c_I <\mathbf{q}> = \sqrt{(s_1+c_Ib_1)(s_2+c_{II}b_2)}(<\mathbf{e}> + \omega_I<\mathbf{h}>) \\ <\mathbf{j}> + c_{II} <\mathbf{q}> = \sqrt{(s_1+c_Ib_1)(s_2+c_{II}b_2)}(<\mathbf{e}> + \omega_{II}<\mathbf{h}>)\end{array}\right\}. \tag{44}$$

Solving these equations with respect to $<\mathbf{j}>$ and $<\mathbf{q}>$, we get all the four EKC.

In the most general form the method of isomorphism was developed in [21] and realized not only for thermoelectric phenomena, but also for two- and three-dimensional galvanomagnetic phenomena and the problem of conductivity of anisotropic media. Afterwards this method was extended to thermogalvanomagnetic phenomena [26]. In [23] material equations (1) are written as

$$\mathbf{j}_a = \sum_b \sigma_{ab}\mathbf{E}_b, \ a,b = 1,2, \tag{45}$$

where $\mathbf{j}_1 \equiv \mathbf{j}$, $\mathbf{j}_2 \equiv \mathbf{q}/T$ and it is shown that by means of linear transformations

$$\mathbf{E}_a = \sum_b \hat{I}_{ab}\mathbf{E}'_b, \ \mathbf{j}_a = \sum_b N_{ab}\mathbf{j}_b, \tag{46}$$

the matrix of local kinetic coefficients $\sigma_{ab}$ can be diagonalized for two phases simultaneously, i.e. in

the hatched system

$$\mathbf{j}_a = \sigma'_b(\mathbf{r})\mathbf{E}'_b. \tag{47}$$

Omitting the details of awkward calculations, we will cite only the final result. Let the solution of the problem on calculating effective electric conductivity $\sigma_{\tilde{a}}$ in a medium without thermoelectric phenomena be known and written as

$$\sigma_e = \sigma_1 f(p,h), \tag{48}$$

where $h = \sigma_2/\sigma_1$ and $p$ – as a rule, concentration of the first (well conducting phase), and the matrix of kinetic coefficients $\hat{\sigma}$ in thermoelectric case has the form

$$\hat{\sigma}_i = \begin{pmatrix} \sigma_i & \gamma_i \\ \gamma_i & \chi_i \end{pmatrix}, \ \gamma_i = \sigma_i \alpha_i, \ \chi_i = \kappa/\dot{O} + \sigma_i \alpha_i^2, \tag{49}$$

where $i$=1, 2 denote phase number.

Then effective kinetic coefficients [23] are of the form:

$$\sigma_{\tilde{a}} = \frac{(\mu\gamma_1 - \gamma_2)f(p,\lambda) - (\lambda\sigma_1 - \sigma_2)f(p,\mu)}{\mu - \lambda}, \tag{50}$$

$$\alpha_{\tilde{a}} = \frac{(\mu\gamma_1 - \gamma_2)f(p,\lambda) - (\lambda\gamma_1 - \gamma_2)f(p,\mu)}{(\mu\sigma_1 - \sigma_2)f(p,\lambda) - (\lambda\sigma_1 - \sigma_2)f(p,\mu)}, \tag{51}$$

$$\kappa_e = \frac{\sigma_1 \kappa_1 (\mu - \lambda) f(p,\lambda) f(p,\mu)}{(\mu\sigma_1 - \sigma_2)f(p,\lambda) - (\lambda\sigma_1 - \sigma_2)f(p,\mu)}, \tag{52}$$

where functions $f(p,\lambda)$ and $f(p,\mu)$ are the same as in (48) (with substitution of $h$ by $\lambda$ and $\mu$, respectively), and $\mu$ and $\lambda$ are equal to:

$$\begin{Bmatrix}\mu\\ \lambda\end{Bmatrix} = \frac{1}{4\sigma_1\kappa_1}\left\{\sqrt{\left(\sqrt{\sigma_1\kappa_1} + \sqrt{\sigma_1\kappa_2}\right)^2 + \sigma_1\sigma_2\dot{O}(\alpha_1 - \alpha_2)^2} \pm \right. \\ \left. \pm \sqrt{\left(\sqrt{\sigma_1\kappa_2} - \sqrt{\sigma_2\kappa_2}\right)^2 + \sigma_1\sigma_2\dot{O}(\alpha_1 - \alpha_2)^2}\right\}. \tag{53}$$

According to method of derivation of these relationships, (50) – (52) are valid for problems of any dimensionality, two-phase media with any structure (arrangement of phases, when $\sigma_e$ in problem (48) is isotropic).

Note that elimination of functions $f(p,\lambda)$ and $f(p,\mu)$ from (50) – (52) gives a relationship connecting $\sigma_e$, $\kappa_e$, $\alpha_e$ and independent of specific medium structure [23]

$$\frac{\kappa_e}{\sigma_e} = \dot{O}\frac{\gamma_1\chi_2 - \gamma_2\chi_1 - (\sigma_1\chi_2 - \sigma_2\chi_1)\alpha_e - (\sigma_2\gamma_1 - \sigma_1\gamma_2)\alpha_e^2}{\sigma_1\gamma_1 - \sigma_2\gamma_2}, \tag{54}$$

In the case when the influence of thermoelectric phenomena on electric and thermal conductivity is low, i.e. with low figure of merit $Z_iT \ll 1$, the relationship (54) changes to a simpler one [17, 23]

$$\alpha_e = \frac{\alpha_1\sigma_1\kappa_2 - \alpha_2\sigma_2\kappa_1 - \sigma_1\sigma_2(\alpha_1-\alpha_2)\kappa_e/\sigma_e}{\sigma_1\kappa_2 - \sigma_2\kappa_1} = $$
$$= \alpha_1 + \frac{\sigma_1\sigma_2(\alpha_1-\alpha_2)}{\sigma_1\kappa_2 - \sigma_2\kappa_1}\cdot\left(\frac{\kappa_1}{\sigma_1} - \frac{\kappa_e}{\sigma_e}\right), \qquad (55)$$

where now $\sigma_e = \sigma_1 f(p,\sigma_2/\sigma_1)$, $\kappa_e = \kappa_1 f(p,\kappa_2/\kappa_1)$.

Thus, behaviour of effective coefficient $\alpha_e$ in critical area $(|\tau|\ll 1)$ is determined by behaviour of effective conductivity (calculated without regard to thermoelectric phenomena), for this purpose it is sufficient to substitute the function $f(p,h)$ from (49) into (50) – (52). In particular, it means that critical indexes describing behaviour of effective thermoelectric coefficient $\alpha_e$ close to flow threshold, can be combinations of only $t$, $q$ and $\nu$ – critical indexes describing behaviour of $\sigma_e$.

Relationship (55) to a great accuracy was checked by numerical simulation [19] and experimentally [27].

As discussed above – within EMT-approximation and by means of hierarchical model of percolation structure – behaviour of $\alpha_e$ at $\kappa_1 \approx \kappa_2$ and $\kappa_1 \gg \kappa_2$ is qualitatively different.

Consider first the case $\kappa_1 \approx \kappa_2$. If the following inequality holds true

$$\left(\frac{\sigma_2}{\sigma_1}\right)^{\frac{t}{t+q}} \ll \frac{\alpha_2\sigma_2}{\alpha_1\sigma_1} \ll 1, \qquad (56)$$

then, as follows from (55), effective coefficient $\alpha_e$ is essentially changed in the critical area

$$\alpha_e \approx \alpha_1, \qquad \delta > \delta_e, \qquad \tau^t \gg \alpha_2\sigma_2/\alpha_1\sigma_1, \ t \gg \Delta, \ \text{I}, \qquad (57)$$

$$\alpha_e = \alpha_2\frac{\sigma_2}{\sigma_1}\tau^{-t}, \qquad p > p_e, \ \Delta \ll \tau \ll \left(\frac{\alpha_2}{\alpha_1}\frac{\sigma_2}{\sigma_1}\right)^{\frac{1}{t}}, \ \text{II}, \qquad (58)$$

$$\alpha_e = \alpha_2\left(\frac{\sigma_2}{\sigma_1}\right)^{\frac{q}{t+q}} = \alpha_2\Delta^q, \ |\tau| < \Delta, \ \text{III}, \qquad (59)$$

$$\alpha_e = \alpha_2|\tau|^q, \qquad p < p_e, \ |\tau| \gg \Delta, \ \text{IV}, \qquad (60)$$

where $(\sigma_2/\sigma_1)^{\frac{t}{t+q}} = \Delta$, as before, stands for smeared area.

As an illustration, regularities (57) – (60) are shown in Fig.5.

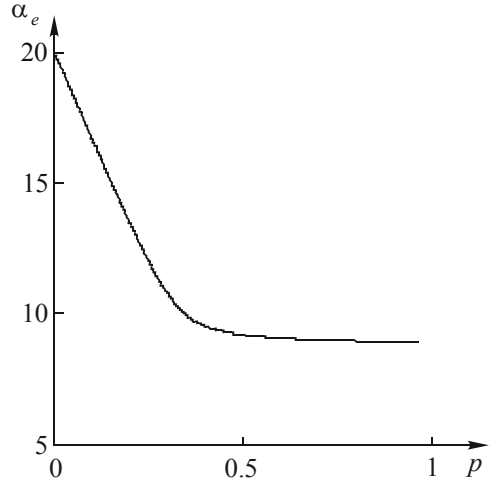

*Fig.5. Schematic of $\alpha_e$ behaviour in critical area at $\kappa_1 \approx \kappa_2$ (see also Fig.1a).*

It is interesting to note that above the flow threshold beyond the smeared area, where the "key role" should be seemingly played by the first phase ($\sigma_1$, $\kappa_1$, $\alpha_1$), in the II area $\alpha_e \approx \alpha_2$.

Otherwise, at $\kappa_1 \gg \kappa_2$ a change in $\alpha_e$ from $\alpha_1$ to $\alpha_2$ occurs in critical area, unlike the case $\kappa_1 \approx \kappa_2$, when $\alpha_e$ varies in the entire area of concentration change (see EMT-approximation, providing good accuracy beyond the critical area). At $\kappa_1 \gg \kappa_2$ several different cases are possible. At $\sigma_2/\sigma_1 \ll \alpha_2\sigma_2/\alpha_1\sigma_1 \ll 1$ or, what is the same, $\alpha_2 \gg \alpha_1$ and $\alpha_2\sigma_2 \ll \alpha_1\sigma_1$, as well as $\sigma_2/\sigma_1 = \kappa_2/\kappa_1$, a mathematical ambiguity emerges in (55). Analysis of general expressions (50)–(52) shows that in this case in a linear with respect to $\alpha_i$ approximation the effective coefficient $\alpha_e$ is expressed not only through function $f(p,h)$ (48), but also through its derivative [22]:

$$\alpha_e = \alpha_1 + (\alpha_2 - \alpha_1)\frac{\partial \ln f(p,h)}{\partial h}, \quad h = \frac{\sigma_2}{\sigma_1} = \frac{\kappa_2}{\kappa_1}. \tag{61}$$

In this case

$$\alpha_e \approx \alpha_1, \quad \tau \gg \left(\frac{\alpha_2}{\alpha_1}\right)^{\frac{1}{t+q}}\Delta, \quad p > p_e, \text{ I}, \tag{62}$$

$$\alpha_e \approx \alpha_2\left(\frac{\sigma_2}{\sigma_1}\right)\tau^{-(t+q)}, \quad \Delta \ll \tau \ll \left(\frac{\alpha_2}{\alpha_1}\right)^{\frac{1}{t+q}}\Delta, \quad p > p_e, \text{ II}, \tag{63}$$

$$\alpha_e = \frac{t}{t+q}\alpha_2, \quad |\tau| < \Delta, \text{ III}, \tag{64}$$

$$\alpha_e = \alpha_2, \quad |\tau| \gg \Delta, \quad p < p_e, \text{ IV}. \tag{65}$$

As an illustration, regularities (61)–(65) are shown in Fig.6.

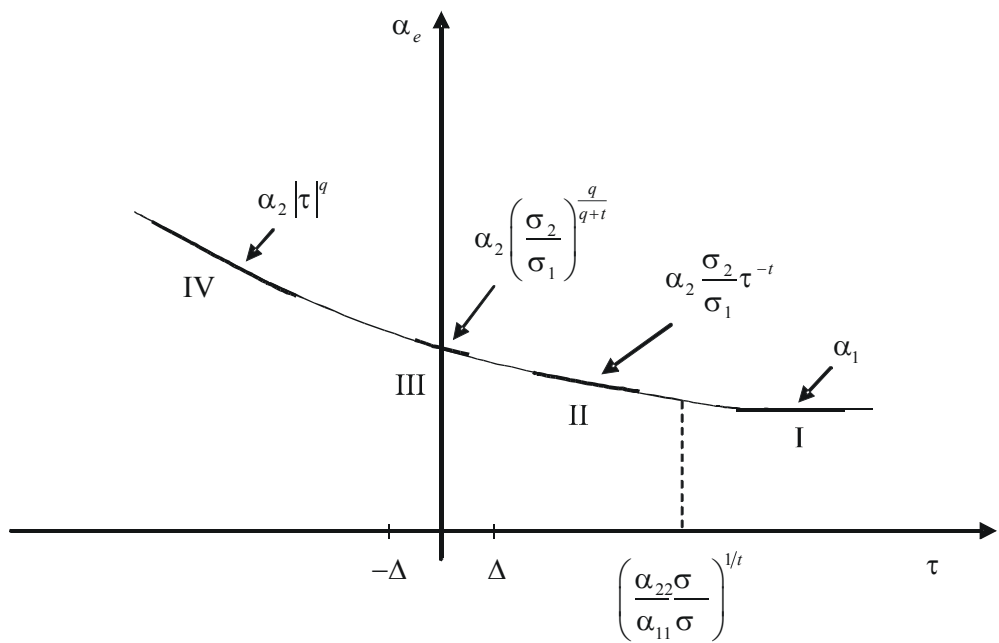

*Fig.6. Schematic of $\alpha_e$ behaviour in critical area at $\alpha_2 \gg \alpha_1$ and $\alpha_2\sigma_2 \ll \alpha_1\sigma_1$, as well as $\sigma_2/\sigma_1 = \kappa_2/\kappa_1$ ( see also Fig.1).*

Considered above was the case $\alpha_2/\alpha_1 \gg 1$, but $(\alpha_2/\alpha_1)(\sigma_2/\sigma_1) \ll 1$, i.e. when inhomogeneity of conductivity $\sigma_2/\sigma_1$ is larger than inhomogeneity of thermoEMF. Otherwise, when $(\alpha_2/\alpha_1)(\sigma_2/\sigma_1) \gg 1$ - inhomogeneity of thermoEMF is dominating, in the smeared area $\alpha_e \approx \alpha_2 \gg 1$. A step close to flow threshold $p_c$ is not so steep now, and output of $\alpha_e$ to «lower» value $\alpha_1$ is delayed, see Fig.7.

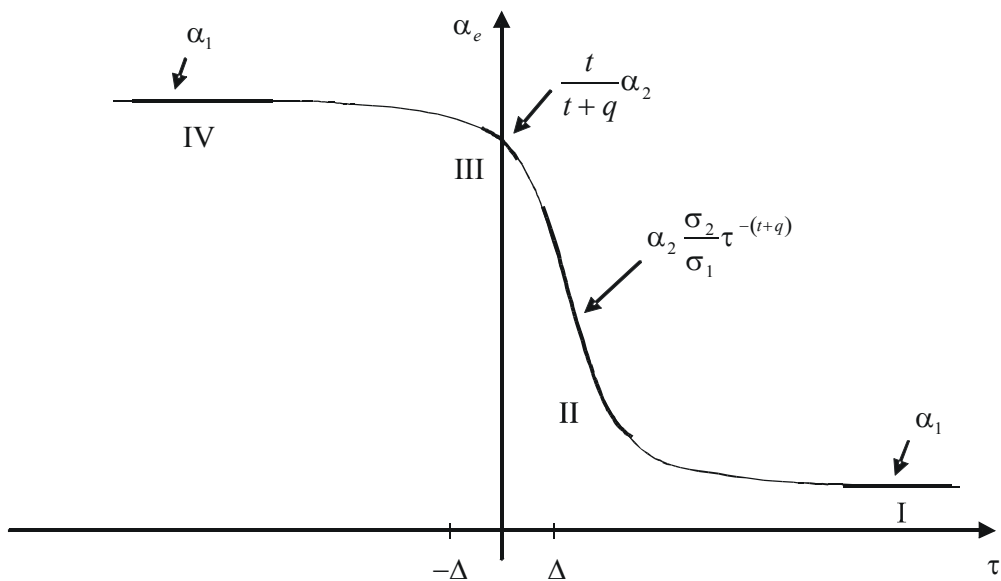

*Fig.7. Schematic of $\alpha_e$ behaviour in the case $\sigma_2/\sigma_1 \neq \kappa_2/\kappa_1$, which is obviously more probable for real materials, behaviour of $\alpha_e$ is similar to that described above (see Fig.6).*

Having described such a powerful technique as isomorphism method, let us come back to the problem of figure of merit $Z_e$ of composite materials. Isomorphism method is applicable when local kinetic coefficients $A_{ij}$ in (13) can be regarded in each of the phases as constants independent of temperature, too. In linear approximation it is surely true, but when calculating the efficiency of thermoelectric devices, and it is for this purpose that $Z_e T$ must be known, one should take into account the influence of thermoelectric phenomena on temperature distribution in the medium, with nonlinearity taken into account as well. Therefore, the problem of $Z_e T$ cannot be considered adequately explored.

In conclusion, it may be said that, of course, the structure of real composites is more complicated than it is the practice in the models discussed above. To take into account this complexity, in some cases it is enough «simply» to add a certain number of contact resistances. In other cases a composite is such a combination of macro- and microinhomogeneities, that the very concept of macro- and microinhomogeneities is poorly determined. This review, in particular, can be helpful in that it makes possible to single out from the properties of complicated inhomogeneous structures the principal, always present, macroscopic regularities, following which one can concentrate on the other, that have not been adequately described yet.

**Conclusions**

The basic regularities of behaviour of effective thermoelectric coefficient of two-phase composites on a physically clear level are explained by the hierarchical model of percolation structure.

The concentration dependence of effective thermoelectric coefficient of two-phase macroscopically inhomogeneous composites in the critical area of concentration is fundamentally dependent on the value of phase thermal conductivity ratio.

For large $ZT$ of phases the effective properties of thermoelectric composites, especially close to flow threshold, invite further investigation.